\documentstyle[onecolumn,mncite]{mn}

\input psfig.sty

\newif\ifAMStwofonts

\newcommand{\grad}{\bmath{\nabla}}
\newcommand{\ve}{\bmath{e}}
\newcommand{\vn}{\bmath{n}}
\newcommand{\vT}{\bmath{T}}
\newcommand{\AU}{\mbox{\small AU}}

\ifoldfss
  \ifCUPmtlplainloaded \else
    \NewTextAlphabet{textbfit} {cmbxti10} {}
    \NewTextAlphabet{textbfss} {cmssbx10} {}
    \NewMathAlphabet{mathbfit} {cmbxti10} {} 
    \NewMathAlphabet{mathbfss} {cmssbx10} {} 
  \fi
  \ifAMStwofonts
    \ifCUPmtlplainloaded \else
      \NewSymbolFont{upmath} {eurm10}
      \NewSymbolFont{AMSa} {msam10}
      \NewMathSymbol{\upi}     {0}{upmath}{19}
      \NewMathSymbol{\umu}     {0}{upmath}{16}
      \NewMathSymbol{\upartial}{0}{upmath}{40}
      \NewMathSymbol{\leqslant}{3}{AMSa}{36}
      \NewMathSymbol{\geqslant}{3}{AMSa}{3E}

       \let\le=\leqslant
       \let\ge=\geqslant
    \fi
  \fi
\fi 

\ifnfssone
  \newmathalphabet{\mathit}
  \addtoversion{normal}{\mathit}{cmr}{m}{it}
  \addtoversion{bold}{\mathit}{cmr}{bx}{it}
  \newmathalphabet{\mathbfit} 
  \addtoversion{normal}{\mathbfit}{cmr}{bx}{it}
  \addtoversion{bold}{\mathbfit}{cmr}{bx}{it}
  \newmathalphabet{\mathbfss} 
  \addtoversion{normal}{\mathbfss}{cmss}{bx}{n}
  \addtoversion{bold}{\mathbfss}{cmss}{bx}{n}
  \ifAMStwofonts
    \ifCUPmtlplainloaded \else
      \UseAMStwoboldmath
      \makeatletter
      \new@mathgroup\upmath@group
      \define@mathgroup\mv@normal\upmath@group{eur}{m}{n}
      \define@mathgroup\mv@bold\upmath@group{eur}{b}{n}
      \edef\UPM{\hexnumber\upmath@group}
      \new@mathgroup\amsa@group
      \define@mathgroup\mv@normal\amsa@group{msa}{m}{n}
      \define@mathgroup\mv@bold\amsa@group{msa}{m}{n}
      \edef\AMSa{\hexnumber\amsa@group}
      \makeatother
      \mathchardef\upi="0\UPM19
      \mathchardef\umu="0\UPM16
      \mathchardef\upartial="0\UPM40
      \mathchardef\leqslant="3\AMSa36
      \mathchardef\geqslant="3\AMSa3E

       \let\le=\leqslant
       \let\ge=\geqslant
    \fi
  \fi
\fi 

\ifnfsstwo
  \DeclareMathAlphabet{\mathbfit}{OT1}{cmr}{bx}{it}
  \SetMathAlphabet\mathbfit{bold}{OT1}{cmr}{bx}{it}
  \DeclareMathAlphabet{\mathbfss}{OT1}{cmss}{bx}{n}
  \SetMathAlphabet\mathbfss{bold}{OT1}{cmss}{bx}{n}
  \ifAMStwofonts
    \ifCUPmtlplainloaded \else
      \DeclareSymbolFont{UPM}{U}{eur}{m}{n}
      \SetSymbolFont{UPM}{bold}{U}{eur}{b}{n}
      \DeclareSymbolFont{AMSa}{U}{msa}{m}{n}
      \DeclareMathSymbol{\upi}{0}{UPM}{"19}
      \DeclareMathSymbol{\umu}{0}{UPM}{"16}
      \DeclareMathSymbol{\upartial}{0}{UPM}{"40}
      \DeclareMathSymbol{\leqslant}{3}{AMSa}{"36}
      \DeclareMathSymbol{\geqslant}{3}{AMSa}{"3E}

       \let\le=\leqslant
       \let\ge=\geqslant
    \fi
  \fi
\fi 

\ifCUPmtlplainloaded \else
  \ifAMStwofonts \else 
    \def\upi{\pi}
    \def\umu{\mu}
    \def\upartial{\partial}
  \fi
\fi

\title{ Nonlinear Single-Armed Spiral Density Waves in Nearly Keplerian Disks}
\author[Erick Lee and Jeremy Goodman]
	{Erick Lee and Jeremy Goodman\\
        Princeton University Observatory, Princeton, NJ, 08544--1001,
	 U.S.A.}

\begin{document}

\maketitle

\label{firstpage}

\begin{abstract}
Single-armed, stationary density waves can exist even in
disks with only weak self-gravity, provided that the rotation curve is
dominated by a central mass.  Such waves could play a significant role
in the transport of angular momentum.  By variational methods, we
derive nonlinear versions of the dispersion relation, angular momentum
flux, and propagation velocity in the tight-winding limit.  The pitch
angle increases with amplitude until the tight-winding approximation
breaks down.  By other methods, we find a series of nonlinear
logarithmic spirals which is exact in the limit of small disk mass and
which extends to large pitch angle.  These waves may be supported by
low-mass protoplanetary disks, and perhaps by compact molecular disks
in galactic nuclei.

\end{abstract}


\section{Introduction}

The number of spiral arms in selfgravitating disks
is influenced by the rotation curve.
As observed by Lindblad,
near-harmonic potentials and linearly rising rotation
curves favor two-armed spirals:
free-particle orbits are slowly precessing ellipses centered
on the minimum of the potential; spiral waves can be built up
from such orbits with relatively weak interactions needed to
force them to precess at a common rate
(cf. \scite{BT}).
Since galactic potentials are often approximately harmonic
at small radii, this construction may in part explain the
prevalence of two-armed spirals and bars in disk galaxies.

The rotation curves of astrophysical disks smaller
than galaxies are usually dominated by a central point-like mass:
examples include accretion disks, protoplanetary disks, planetary
rings, and masing molecular
disks recently discovered in galactic nuclei (Section 4 and 
references therein).
Free-particle orbits in such disks are slowly precessing keplerian
ellipses.  In this case,
Lindblad's construction favors single-armed spirals,
again with small angular pattern speed.
Given the astrophysical importance of keplerian
disks, it is unfortunate that single-armed spirals have received
much less theoretical attention than double-armed ones.
The visual fascination of beautiful spirals in
galactic disks, and the absence of comparably well resolved images
of keplerian disks, may partly explain the neglect.

The present paper is therefore dedicated to the propagation of 
single-armed spirals in nearly keplerian gaseous disks.
We concentrate on the limit where the self-interaction of the disk is weak
compared to its interaction with the central mass.  This implies
small or vanishing angular pattern speeds.
The nonlinear regime is emphasized because it is gratifyingly
tractable, and more importantly because one would like to know the
maximum angular momentum flux that can be transmitted by density waves.
We do not study the linear instabilities of single-armed waves
(cf. \pcite{ARS}; \pcite{STAR}; \pcite{HPS}) or their excitation by companion
masses \cite{YCa85}.

In  Section 2, we adopt the familiar tight-winding approximation
and extend it to nonlinear spirals.
This has already been done using other methods, notably by
\scite{SYL}; \scite{SDLYC}; and 
Borderies, Goldreich, \& Tremaine (1985, 1986).
Our mathematical approach, however, is adapted from the variational
methods of \scite{Whitham}, which are particularly efficient for
describing the propagation of a nonlinear wave train outside the region
where it is excited, especially if dissipation is weak.
In addition to the nonlinear dispersion relation and angular momentum
flux, the variational approach provides the nonlinear analog of the group
velocity, which as far as we know have not otherwise
been derived for density waves (Appendix B).
Also, previous work has mainly been concerned with nonlinear extensions
of the long branch of the WKB dispersion relation, whereas we
are primarily concerned with the short branch.

We show in Section 2 that for single-armed waves in keplerian potentials,
self-gravity is best measured not
by the familiar $Q$ parameter \cite{Toomre}, but by
the ratio (\ref{sigmadef}) of ``Jeans length''
($\lambda_J$) to disk radius.  In fact, almost all of our analyses
pertain to the limit as $Q\to\infty$ at finite $\lambda_J/r$.

The results of  Section 2 show that single-armed waves
unwind with increasing nonlinearity.
Therefore  Section 3 presents nonlinear logarithmic spirals that do not
depend upon the tight-winding approximation.
The method of their construction assumes radial self-similarity,
which restricts us to disks whose azimuthally averaged surface
density scales with radius as $r^{-3/2}$.
We believe, however, that many of the results are more generally
valid.  Solutions do not seem to exist unless
$\lambda_J/r\la 4$.

We summarize our results in Section 4 and discuss applications to
astrophysical disks.  In particular, we show that the minimum-mass
solar nebula was probably in the regime considered here, namely large
$Q$ but moderate $\lambda_J/r$, so that stationary single-armed waves
may have been important.

\section {Tightly-Wrapped Waves}
In this section, we consider single-armed waves of small pitch
angle.
In  Section  2.1, we summarize the linear theory, emphasizing
those aspects that are peculiar to $m=1$ waves in nearly keplerian
potentials.
These results provide background and guidance for the development
of the nonlinear theory in  Section  2.2, which itself has two parts:
first we sketch a variational formalism for nonlinear wave trains in
general, and then we apply this formalism to tightly-wrapped
spiral density waves.

\subsection{Linear Theory}
The dispersion relation for tightly wrapped $m=1$ spiral waves is 
\begin {equation}\label{WKBdisp}
(\omega-\Omega)^2 = \kappa ^2 - 2 \pi G \Sigma k + c^2 k^2.
\label{eq:linWKBdisp}
\end {equation}
However it is excited, 
as the wave propagates inward, the pattern speed ($\omega$)
becomes negligible when compared to the local angular velocity
of the disk ($\Omega$).  We study the stationary limit, $\omega\ll\Omega$,
which is appropriate to waves far inside their corotation radius
close to a central point mass.  In this region, both
$\kappa^2$ and $\Omega^2$ are dominated by contributions from the
central mass, but these large terms cancel one another to leading
order in equation (\ref{eq:linWKBdisp}), leaving a residual due only to
the disk.

To simplify the analysis we consider a razor thin disk and take a polytropic
equation of state $P=K \Sigma^\gamma$.  For a logarithmic spiral pattern,
the density and temperature profiles should be power laws, $\Sigma \propto
r^{-\beta-1}$ and $ c^2 \propto r^{-\beta}$ where $ 0< \beta < 1$.  This
requires  $\gamma = (1+2\beta)/(1+\beta)$, so $\gamma$
can range from 1 to $3/2$.
Now $\kappa^2-\Omega^2$, the self-gravity term, and the pressure term in the
dispersion relation all have the same scaling with $r$ if the pitch angle
$\cot^{-1}(kr)\approx 1/kr$ is constant.
Introducing the dimensionless parameters
\begin{equation}\label{sigmadef}
\sigma^2 \equiv \frac{c^2}{ 2 \pi G \Sigma r } =\frac{\lambda_J}{2\pi r},
\end{equation}
and
\begin{equation}\label{gdef}
g(\beta) \equiv \frac{(\kappa^2 - \Omega^2) r}{\pi G \Sigma}
~~\approx  2~+0.7536\beta(1-\beta)-2\sigma^2(1-\beta^2)
~~~~\mbox{if}~~0\le\beta\le 1.
\end{equation}
we write the dispersion relation in dimensionless form
\begin {equation}\label{lindisp}
g-2|k|r+2\sigma^2 (kr)^2=0.
\end {equation}
Notice that $\kappa^2-\Omega^2$ is independent of the central point mass.
The term involving $\sigma^2$ in eq.~(\ref{gdef}) expresses the influence
of pressure on the disk rotation curve; the rest is due to the disk's
self-gravity.
For small values of $\sigma^2$, the pitch angle 
$\cot^{-1}(kr)\approx\sigma^2$
for the physically relevant root of the dispersion relation (see Section 3.1),
so the $\kappa^2-\Omega^2$ term is negligible.

To see how the amplitude of the linear wave varies with radius we use the fact
that wave angular momentum is conserved.  The angular momentum flux
due to spiral waves arises from gravitational and pressure torques, and from 
advective transport.
The advective transport is proportional to $(1-c^2|k|/2 \pi G \Sigma)$
\cite{GT} which vanishes to leading order in $1/kr$ for these waves.
The flux due to gravitational and pressure torques
can be calculated for tightly-wound patterns in the WKBJ approximation:
\begin {equation}
\Gamma_{\scriptscriptstyle WKB} = \mbox{sign}(k)
 \frac { r \delta \Phi^2} {4 G}=
\mbox{sign}(k)\frac{\pi^2 rG\delta\Sigma^2}{k^2}
\label{eq:linflux}
\end {equation}
Hence $d\Gamma_{\scriptscriptstyle WKB}/dr=0$ implies 
$\delta \Sigma / \Sigma \propto r^{\beta-(1/2)}$, so that
the fractional amplitude of the wave increases inward if $\beta <1/2$.

The $m=1$ waves have another special feature worth remarking upon here:
in principle, they may exert a net force on the central mass.
The component of the force along direction $\theta$ in
polar coordinates is
\begin{displaymath}
\int\limits_0^{2\pi}\int\limits_0^\infty\frac{G\delta\Sigma(r,\theta')}
{r^2}\cos(\theta-\theta')\,rdrd\theta.
\end{displaymath}
If $\delta\Sigma(r,\theta')\propto\exp(im\theta')$, then the integral
above can be nonzero for $m=1$.
In all cases of interest to us, however, $\delta\Sigma$ will be
an oscillatory (wavelike) function of radius, so that the
force is dominated by the endpoints of the radial integration.
Thus, the displacement of the central mass cannot be included in
a local analysis such as ours, whether linear or nonlinear, and
we shall neglect it.  For a careful global treatment of this
issue in nonlinear as well as linear regimes, see \scite{HPS}.
These authors find that modes capable of displacing the central
mass tend to be evanescent in radius rather than wavelike.

\subsection{Nonlinear Theory}
Approximate solutions for a wide variety of nonlinear wave trains arising
from mathematical physics can be found with remarkable
efficiency by varational methods.
Since these methods are not widely used in astrophysics,
we digress briefly to summarize the main ideas.
For a more complete exposition, see \scite{Whitham}.

The variational approach rests on the following assumptions:
\begin{itemize}

\item[(A.1)] The fundamental equations of motion
are derivable from an action principle, hence generally
nondissipative.

\item[(A.2)] The wavelength and wave period of interest
are small compared to the length scale and time scale
of variations in the background.

\item[(A.3)] The waves are locally periodic in time and space, at
least approximately.

\end{itemize}

Assumption (A.1) can be relaxed by tacking dissipative terms onto the
equations of motion at a late stage in the procedure, but we will not
do so here.  Assumptions (A.2) and (A.3) are shared with linear WKBJ
theory, but since the principle of superposition does not
hold, one cannot synthesize a solitary wave---a highly localized
wave packet---by linearly combining periodic wave trains.  On the other
hand, the third assumption permits gradual changes in the wavelength,
wave period, and wave amplitude, provided that they occur on scales
long compared to the wavelength and period.

With these assumptions, the following procedure accurately 
approximates nonlinear wave trains.
The dynamical variable ($u$) depends in the first instance
on one or more independent spatial variables ($x_1,\ldots,x_n$) and
on time ($t$); $u$ might in fact represent multiple  dependent variables.
By assumption (A.1), $u$ obeys an action principle:
\begin{equation}\label{action_principle}
\delta\int L(u,\partial_t u,\partial_1 u,\ldots,\partial_n u;x_1,\ldots,
x_n,t){\cal J}(x)d^nx dt=0.
\end{equation}
Here ${\cal J}(x)$ is just the Jacobian of the transformation from
$(x_1,\ldots,x_n)$ to cartesian coordinates.
We have indicated that $L$ depends on first spatial derivatives of
$u$, but in fact it may depend on spatial derivatives of any order,
as is the case for self-gravitating density waves.
For solutions of the action principle obeying assumption (A.3), $u$ 
can be written as a periodic function
of a single phase ($\Psi$):
\begin{equation}\label{wave_ansatz}
u = U(\Psi)=U(\Psi+2\pi),~~~~
\Psi = \Psi(x_1,\ldots,x_n,t).
\end{equation}
The frequency and wavenumber are defined as derivatives of $\Psi$: 
\begin{equation}\label{dPsidef}
\omega\equiv-\partial_t\Psi,~~~~
k_i\equiv\partial_i\Psi\equiv\frac{\partial\Psi}{\partial x_i}.
\end{equation}
Hence $\partial_t u\to-\omega U'$ and $\partial_i u\to k_i U'$.
The critical approximation is now to \emph{average over the phase
$\Psi$ while pretending that $\omega,k_1,\ldots,k_n$ and
$t,x_1,\ldots x_n$ are constant}:
\begin{equation}\label{phaseaverage}
\bar L(U,U',\omega,k_1,\ldots;x_1,\ldots,t)\equiv
\int\limits_0^{2\pi} L(U,-\omega U',k_1 U',\ldots;x_1,\ldots,t)
{\cal J}(x)\frac{d\Psi}{2\pi}.
\end{equation}
The action principle (\ref{action_principle}) reduces to
\begin{equation}\label{average_action}
\delta\int \bar L(U,U';\omega,k_1,\ldots;x_1,\ldots,t)
{\cal J}(x)d^nx dt=0,
\end{equation}
where $\delta$ now indicates variation of $U(\Psi)$ and 
$\Psi(x_1,\ldots,x_n,t)$.

Stationarity of the action with respect to $\Psi$,
which enters only through its derivatives (\ref{dPsidef}),
implies the conservation law
\begin{equation}\label{Euler_Lagrange}
\frac{\partial}{\partial t}\left({\cal J}
\frac{\partial{\bar L}}{\partial\omega}
\right)-\sum\limits_{i=1}^n\frac{\partial}{\partial x_i}\left({\cal J}
\frac{\partial{\bar L}}{\partial k_i}\right)=0.
\end{equation}
The quantity $\partial\bar L/\partial\omega$
is the density of wave action, and
$-\partial\bar L/\partial k_i$ is the $i^{\rm th}$
component of the corresponding flux.

Stationarity of the action with respect to $U$ yields both
the nonlinear dispersion relation and also equations for the functional
form of $U(\Psi)$ [cf. \pcite{Whitham}].

To apply this formalism to spiral density waves, the radial wavelength
must be small compared to the radius in the disk ($r$), since the surface
density and other background properties vary significantly over radial
distances comparable to $r$.
For single-armed spirals, a short radial wavelength implies that the wave 
is tightly wrapped.
So we may as well take advantage of tight winding
to simplify our formulae.

For dynamical variables, we take the radial and angular displacements
of a fluid particle from a circular orbit:
\begin {equation}
\delta r\equiv r_{\rm p}(t),~~~~
\delta\theta = \theta_{\rm p}(t)-\Omega(a)t-\theta_{\rm p}(0),
\end {equation}
where $[r_{\rm p}(t),\theta_{\rm p}(t)]$ are the polar coordinates
of the particle at time $t$.
Here $a$ and $\theta_p(0)$ are labels that follow the particle, but
we will soon interpret $a$
as the semimajor axis of a keplerian orbit that closely approximates
the particle's trajectory.  $\Omega(a)$ is the mean angular velocity of
that trajectory, which is not exactly the same as the mean motion of
a keplerian orbit of semimajor axis $a$, because the pressure and
self-gravity of the disk modify its rotation curve.
For independent variables we take $(a,\theta,t)$.
In tightly-wound waves where the
streamlines do not actually cross, $\delta r/a\ll 1$ but
$\partial\delta r/\partial a$ may approach unity.
Therefore in constructing the lagrangian, we keep only the 
lowest important order in $\delta r/a$ but all orders in
$\partial\delta r/\partial a$.

In the absence of any collective effects, the lagrangian and the
action would be those of a collection of free particles:
\begin{equation}\label{L_free}
I_{\rm free}= \int\limits_{t_1}^{t_2}\int\limits_0^{2\pi}
\int\limits_0^\infty
L_{\rm free}(\delta r,\delta\dot r,a)\,ada d\theta dt,~~~~~~
L_{\rm free}=
\frac{1}{2}\left[\left(\frac{d\delta r}{dt}\right)^2
-\kappa^2(\delta r)^2\right] \Sigma_0(a),
\end{equation}
where
\begin{equation}\label{kappa_def}
\kappa^2\equiv \frac{1}{2a^3}\frac{d}{da}\left(a^4\Omega^2\right)
\end{equation}
is the epicyclic frequency, and $\Sigma_0(a)$ is the surface
density that fluid elements would have in the unperturbed
axisymmetric disk.
The corresponding Euler-Lagrange equation for each fluid element is
$\delta\ddot r=-\kappa^2 r$.

To represent the self-interaction of the disk, we
subtract from $I_{\rm free}$ the time integral of
two potential-energy terms.
The first of these is the thermodynamic internal energy,
$W_{\rm int}$.
Absent dissipation,
the two-dimensional pressure and density of a given fluid
element scale as $P\propto\Sigma^\gamma$, and
the internal energy per unit mass is
$\gamma P/(\gamma-1)\Sigma$.
Let $P_0(a)$ be the 2D pressure that a
fluid element would have in a circular orbit.
Its actual surface density is
\begin{equation}\label{Sigma}
\Sigma(a,\theta,t)= \Sigma_0(a)\left(1+\frac{\partial\delta r}
{\partial a}\right)^{-1}
\end{equation}
in the tight-winding approximation.
It follows that the total internal energy of the gas is
\begin{equation}\label{W_int}
W_{\rm int}= \int\int\frac {c^2_0(a)}{\gamma(\gamma -1)}
\left[ \left( 1+\frac {\partial \delta r} {\partial a} \right) 
^{-(\gamma-1)} -1 \right]\Sigma_0(a) a da d\theta,
\end {equation}
where $c^2_0(a)\equiv \gamma P_0/\Sigma_0$ is the square of the
unperturbed sound speed.

The gravitational self-energy of a tightly-wound wave is accurately
expressed by the ``thin-wire'' approximation.
On the scale of the radial wavelength, curves of constant
surface density are approximately straight lines.
The gravitational interaction energy of two lines of mass
per unit length $(\mu_1,\mu_2)$ and separation $s_{12}$ is
$G \mu_1\mu_2\ln|s_{12}|$ per unit length.
We put $\mu_1=\Sigma_0(a)da_1$, $\mu_2=\Sigma_0(a)da_2$, where
$a\equiv(a_1+a_2)/2$, since $\Sigma_0(a)$ varies only on the scale
of $a$.
And since the pitch angle of the streamlines is small,
$s_{12}=|a_1+\delta r_1-(a_2+\delta r_2)|$, where $\delta r_i\equiv
r(a_i,\theta,t)$ and $\theta\equiv(\theta_1+\theta_2)/2$.
Therefore,
\begin{equation}\label{W_grav}
W_{\rm grav} = G\int\limits_0^{2\pi}d\theta
\int\limits_0^\infty a\Sigma_0^2(a) da
\int\limits_{-2a}^{2a}d(a_1-a_2)
~\ln\left|1+\frac{\delta r_1-\delta r_2}
{a_1-a_2}\right|.
\end{equation}
We have replaced $s_{12}$ by $s_{12}/(a_1-a_2)$ inside the logarithm;
this changes $W_{\rm grav}$ by a term independent of $\delta r$, which
has no effect on the equations of motion.
The integration over $(a_1-a_2)$ is also insensitive to its limits,
since the dominant contribution comes from $|a_1-a_2|\ll a$ in a
tightly-wound wave.
The above expression for $W_{\rm grav}$ describes local self-gravity
but not the long-range contribution
of the unperturbed axisymmetric disk surface density to the rotation
curve.
The latter is included in the mean angular velocity $\Omega(a)$ and
in the epicyclic frequency (\ref{kappa_def}) derived from it.

For single-armed wavelike solutions, $\delta r$ and $\delta\theta$
depend on $(a,\theta,t)$ \emph{via} the phase variable
\begin{equation}\label{kepler_phase}
\Psi\equiv \theta-\phi(a)-\omega t.
\end{equation}
In principle the wavenumber has two components, but since the
angular component
$k_\theta\equiv\partial\Psi/\partial\theta$ is always unity,
we write
\begin{equation}
k\equiv\frac{\partial\Psi}{\partial a}= -\phi'(a)
\end{equation}
for the radial component.

Our next step is to average the full lagrangian
over one period in $\Psi$.
This task is simplified by the fact that we are interested in
situations where the interior mass of the disk
is small compared to that of the central object ($M$).
We introduce a small parameter
\begin{equation}\label{eta_def}
\eta\equiv \frac{2\pi \bar a^2\Sigma_0(\bar a)}{M}\ll 1,
\end{equation}
where $\bar a$ is typical of the radii of interest.
The self-interaction terms $W_{\rm int}$ and $W_{\rm grav}$
are formally of order $\eta$ compared to the free action (\ref{L_free}).
To leading order in $\eta$, the trajectory of a fluid
element must therefore be a solution of the Euler-Lagrange equation
derived from $L_{\rm free}$ alone.
Such a solutions is sinusoidal in $t$, and since it must depend on $t$
through $\Psi$, it can be written in the form
\begin {equation}\label{delta_r}
\delta r(a,\theta)= a e(a) \cos \Psi,
\end {equation}
where $e$ is a dimensionless measure of the amplitude.
For $\eta=0$, the only forces would be those
of an exactly keplerian potential; in that case, $\omega=0$, and
$e$ and $\phi$ could be chosen as arbitrary constants independently
for each fluid element.
For $\eta\ne 0$ but small, 
$e$ and $\phi$ must be smooth functions
of $a$ because the self-interaction terms would
diverge if neighboring streamlines were to intersect.
By substituting the form (\ref{delta_r}) into the lagrangian and applying
the phase-averaged action principle, one finds relations between 
$e$, $\omega$, and $k$ to leading order in $\eta$.
At higher orders in $\eta$, there are corrections to these
relations and to the functional form (\ref{delta_r}).

Note that the small parameter $\eta$ is independent of the
degree of nonlinearity of the wave.
The latter is measured in the tight-winding approximation by
the streamline-crossing parameter
\begin {equation}\label{q0def}
q_0=|k|ae
\end {equation}
Streamlines intersect if and only if $q_0\ge 1$,
as can be seen by differentiating eq.~(\ref{delta_r})
with respect to $a$ assuming $d\ln e/da= O(1/a)\ll k$, and then
referring to eq.~(\ref{Sigma}).
We will assume that $q_0<1$, but not necessarily $q_0\ll 1$,
since otherwise dissipation is inevitable.  Hence since $ka\gg 1$
is necessary for the tightwinding approximation, we must have
$e\ll 1$.

The time derivative in eq.~(\ref{L_free})
is the total time derivative following a
fluid element,
\begin{displaymath}
\frac{d\delta r}{dt}\approx \frac{\partial\delta r}{\partial t} +
\Omega(a)\frac{\partial\delta r}{\partial\theta}.
\end{displaymath}
The angular velocity $\dot\theta=\Omega+O(e)$,
but we ignore the $O(e)$ correction when it multiplies angular derivatives,
since these are small compared to radial derivatives.
Hence the phase average of $L_{\rm free}$ is
\begin{equation}
\int\limits_0^{2\pi} L_{\rm free}\frac{d\Psi}{2\pi}= \frac{\Sigma_0(a)}{4}
\left[(\Omega-\omega)^2-\kappa^2\right]e^2(a).
\end{equation}

The phase averages of the internal and
gravitational energies are derived in Appendix A, and are
expressed in terms of the following integral functions:
\begin {equation}\label{Udef}
{\cal U}(q_0) \equiv \frac {4} {\gamma (\gamma -1)}  q_0^{-2}
\int_0^{2\pi} \frac {d\Psi} {2\pi} \left[ \left (1-q_0 \sin \Psi \right)
^{-(\gamma-1)}-1 \right],
\end {equation}
and
\begin {equation}\label{Wdef}
{\cal W}(q_0) ~\equiv~ - \frac {4} {\pi |q_0|} \int_{-\infty}^{\infty} dy
\ln\left|\frac 1 2 + \frac 1 2 \sqrt{1-\left(\frac{\sin(q_0y)}y\right)^2}
\right|.
\end {equation}
These functions are normalized so ${\cal U} = 1 + {\cal O} (q_0^2)$
and ${\cal W} = 1 + {\cal O}(q_0^2)$.  Both
increase monotonically with $q_0$ and both are finite but nondifferentiable
at $q_0=1$:
${\cal U}^\prime(q_0) \propto (1-q_0)^{1/2-\gamma}$ as $q_0 \rightarrow 1$,
while ${\cal W}^\prime(q_0) \propto \ln (1-q_0)$.

The phase-averaged action is therefore
\begin {equation}\label{Iavg}
I = \int\int  \bar L(e,w,k,a) 2\pi adadt.
\end {equation}
It is convenient to divide $\bar L$ into a part $\bar L_0$ that
is independent of $\omega$ and a residual:
\begin {eqnarray}\label{Lbar}
\bar L(e,\omega,k,a) &=& \bar L_0(e,k,a)
~+~\frac{\Sigma_0(a)}{4}\,\omega[\omega-2\Omega(a)]a^2e^2,\nonumber\\
\bar L_0(e,k,a)&=&
-\frac {\pi G\Sigma_0^2 a}{2} \left[\frac 1 2 g e^2 
- |q_0| {\cal W}(q_0) e + \sigma^2 q_0^2 {\cal U}(q_0) \right]
\end {eqnarray}

\begin{figure}
\psfig{figure=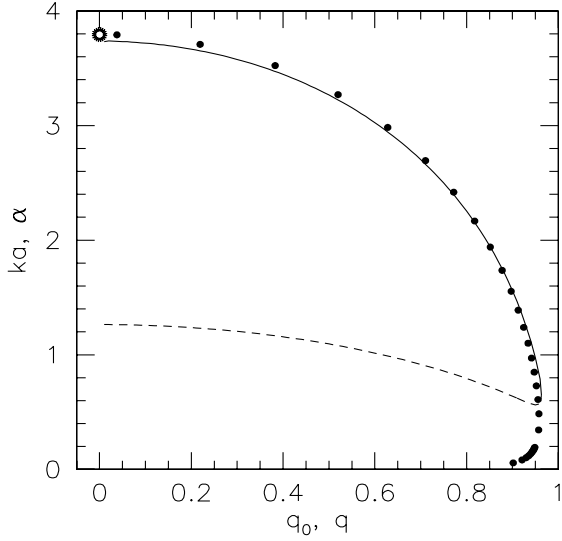,height=8.0truecm,width=8.4truecm}

\caption{Nonlinear dispersion relation for stationary single-armed waves
at $\sigma^2=0.2$, $\beta=1/2$, and $\gamma=4/3$. 
\emph{Abscissa}: streamline-crossing parameter $q_0$ (eq.~[\ref{q0def}])
or $q$ (eq.~[\ref{qdef}]).  \emph{Ordinate}: dimensionless radial
wavenumber $ka$ or $\alpha$, equal to cotangent of pitch angle.
\emph{Solid curve:} Short branch in tight-winding theory ($ka$ vs. $q_0$).
\emph{Dashed curve:} Unphysical long branch.
\emph{Filled Points:} Logarithmic spirals ($\alpha$ vs. $q$).
\emph{Hollow Point:} Linear theory ( Section 3.1)
} \label{dispfig}
\end{figure}

The averaged action (\ref{Iavg}) must be stationary with respect to
$e$; imposing $\partial\bar L/\partial e=0$ yields
\begin {equation}
\label {DISP}
g e^2 - e\,\mbox{sign}(k) \frac {d} {dq_0} [q_0^2{\cal W}(q_0)]
+ \sigma ^2 q_0 \frac {d} {dq_0} [q_0^2{\cal U}(q_0)]
-\omega(\omega-2\Omega)\frac{ae^2}{\pi G\Sigma_0}=0.
\end {equation}
In the limit $e\ll 1$ at fixed $ka$, every term in this expression
is proportional to $e^2$, and we recover the
linear dispersion relation (\ref{WKBdisp}) or (\ref{lindisp}),
since ${\cal U}(0)={\cal W}(0)=1$ and ${\cal U}'(0)={\cal W}'(0)=1$.
In general, however, the dispersion relation relates the amplitude
of the wave to its wavelength and frequency.
If we fix one of these three quantities, then the other two are
constrained to a curve.
In this paper, we are most interested in $\omega=0$.
Figure~\ref{dispfig} 
displays some representative dispersion curves for stationary waves.
We have used $q_0$ rather than $e$ as a measure of wave amplitude because
it more directly controls the importance of
pressure and gravitational forces.

The Euler-Lagrange equation for $\Psi$ is (\ref{Euler_Lagrange}),
with the jacobian ${\cal J}=2\pi a$ in this case.
Hence the wave action density (action per unit physical area) is
\begin{equation}\label{DENSITY}
\rho\equiv \frac{\partial\bar L}{\partial\omega}=
(\omega-\Omega)\frac{\Sigma_0a^2e^2}{2}.
\end{equation}
This can also be interpreted as the angular momentum density of
the wave.  The density is negative for stationary waves ($\omega\to 0$).
The corresponding radial flux is
\begin{eqnarray}\label{FLUX}
f\equiv -\frac{\partial\bar L}{\partial k}&=&\frac{\pi G\Sigma_0^2 a^2 e}{2}
\frac{\partial}{\partial q_0}\left[\sigma^2q_0^2{\cal U}(q_0)
-e|q_0|{\cal W}(q_0)\right] \nonumber\\
&=&\frac{\pi G\Sigma_0^2 a^2e^2}{2}\left[\mbox{sign}(k){\cal W}(q_0)
-\frac{g}{ka}\right]
~+\omega(\omega-2\Omega)\frac{\Sigma_0 a^2e^2}{k}.
\end{eqnarray}
In the second line, we have used the dispersion relation (\ref{DISP})
to eliminate the derivatives of ${\cal U}$ and ${\cal W}$.
Strictly, the term $g/ka$ is negligible in the tightly-wound limit.
From eqs.~(\ref{Sigma})
and (\ref{delta_r}), the fractional surface density fluctuation
becomes $\delta \Sigma / \Sigma = q_0 = kae$ for $q_0\ll 1$.
Therefore in the linear (and stationary and tightly-wound) limit,
$2\pi a f$ reduces to $\Gamma_{\scriptscriptstyle WKB}$  
(eq.~[\ref{eq:linflux}]).

\begin{figure}
\psfig{figure=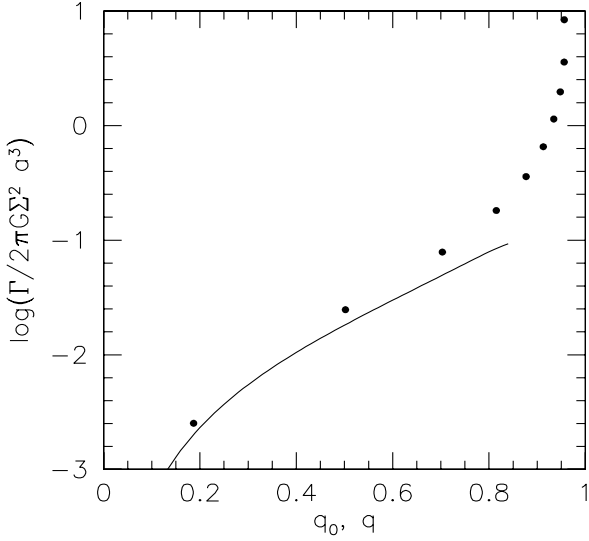,height=8.0truecm,width=8.4truecm}
\caption{Wave torque (azimuthally integrated radial angular momentum flux)
for $\sigma^2=0.2$, $\beta=1/2$, $\gamma=4/3$.
\emph{Solid curve:} Tight-winding theory; $\Gamma=2\pi a f$,  $f$
from eq.~(\ref{FLUX}) with $\omega=0$.
\emph{Points:} Torque of logarithmic spirals ( Section 3.3 and Appendix C).
}\label{fluxfig}
\end{figure}

For stationary waves, $\Gamma\equiv 2\pi a f$ is independent of
radius.  This quantity contains the grouping of disk parameters
$\Sigma^2 a^3$, which scales with radius as $a^{1-2\beta}$.
Therefore if $\beta=1/2$, a stationary wave occupies
a single point $(q_0,ka)$ on a dispersion curve such as those in Fig.~1
at all radii.
If $\beta\ne 1/2$, a radially-propagating wave train
moves along the dispersion curve 
to compensate for the changes in the disk.
This defines $q_0$ and $e$ as functions of $a$, with the angular
momentum flux $\Gamma$ as a parameter.
In particular, if $\beta < 1/2$, tightly-wound trailing waves tend to
become more nonlinear as they propagate inward,
because $\Gamma/(\Sigma^2 a^3)$ tends to increase with $q_0$,
as is shown for $\sigma^2=0.2$ in Figure~\ref{fluxfig}.
Unfortunately, as Figure~\ref{dispfig} demonstrates, the pitch angle
$\cot^{-1}(ka)$ also increases with $q_0$ until the tight-winding
approximation can no longer be trusted.
This generally happens well before the point of streamline crossing
is reached ($q_0=1$).

The stationary dispersion relation (\ref{DISP}) can
be regarded as a quadratic equation in $e$ with coefficients in $q_0$.
The two solutions are analogous to the short and long branches of
linear theory.  Since $k=q_0/ae$, the smaller solution for $e$ is the
short branch. The discriminant is
\begin {equation}
D(q_0) = \left[\frac d {dq_0} (q_0^2 {\cal W}) \right]^2 - 
	4 g \sigma^2 q_0 \frac d {dq_0}(q_0^2 {\cal U})
\end {equation}
The eccentricity $e$ is real only if $D(q_0) >0$.  Since the derivative
of $\cal U$ diverges faster than the derivative of $\cal W$ as
$q_0\to 1$, there is some $q_0<1$ beyond which no real solutions for
the eccentricity exist.  At this critical $q_0$, the
short and long branches join.
These conclusions are academic, however, since the join occurs
outside the tightly-wound regime.
In fact, the long branch does not actually
exist for stationary waves ( Section 3).

We have tacitly assumed that short-branch
stationary trailing waves always propagate inwards.
To verify this assumption for our nonlinear wave trains,
we must investigate the nonlinear generalization of the group velocity.
An obvious candidate is the ratio $f/\rho$ of wave flux to wave density,
which reduces to the radial group velocity in the linear limit.
Nonlinearly, however, there are actually two distinct characteristic
velocities at which information about the wave propagates; neither 
coincides with $f/\rho$  or $v_{\rm group}$ except as $q_0\to0$.
As shown by \scite{Whitham}, this is a generic
property of nonlinear wave trains.
The characteristic velocities in the present case are derived
in Appendix B and exemplified in Figure~\ref{velfig}.
Both characteristic velocities are negative (inward) for trailing
waves, at least as long as the wave remains tightly wound.
\begin{figure}
\psfig{figure=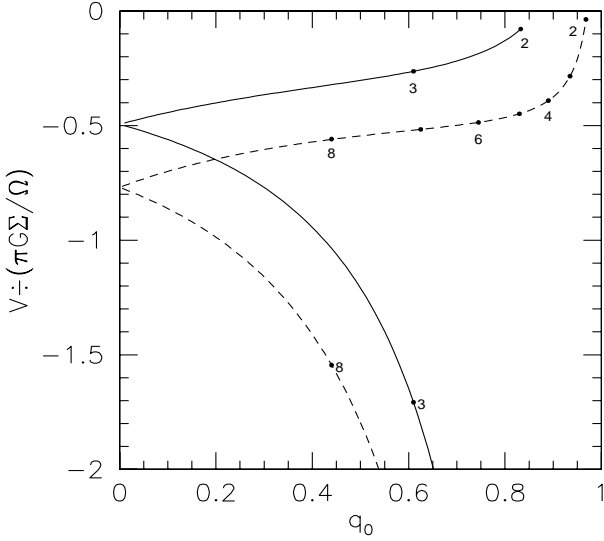,height=8.0truecm,width=8.4truecm}
\caption{
The two characteristic velocities (radial propagation velocities) of
stationary, trailing, single-armed spirals in the tight-winding 
approximation, for $\beta=1/2$, $\gamma=4/3$, $\sigma^2=0.2$ 
(\emph{solid curves}) and $\sigma^2=0.1$ (\emph{dashed}).
The two velocities converge to the linear group velocity as 
the streamline-crossing parameter $q_0\to0$.
Labeled points mark radial
wavenumber $ka$ along each sequence. Smaller $\sigma^2$
makes for tighter winding (larger $ka$) at a given $q_0$.
}\label{velfig}
\end{figure}
\section {Exact self-similar spirals}

The nonlinear wave-action formalism of  Section 2 is flexible but
assumes tightly-wrapped waves, $|kr|\gg 1$.
In this section, we sacrifice generality for accuracy and find solutions
in the form of exact logarithmic spirals of arbitrary pitch angle.

The surface density is
\begin{equation}
\Sigma(r,\theta)= r^{-3/2}S(\theta+\alpha\ln r),
\label{eq:spiraldef}
\end{equation}
where $S$ is periodic [$S(\psi+2\pi)=S(\psi)$] but may be
nonsinusoidal.
The cotangent of the pitch angle of the spiral, $\alpha$,
corresponds to the previous $ka$.
As suggested by the discussion of flux conservation in section  Section 2,
the surface density must vary as $r^{-3/2}$ on average
if the spiral is to have the same strength $\delta\Sigma/\Sigma$ at all radii.
Self-similarity gives the same requirements as in  Section 2; that the
sound speed vary as $r^{-1/2}$, and the adiabatic index must be
$\gamma=4/3$.

\scite{ST96} have studied self-similar nonaxisymmetric power-law disks.
Their solutions omit the central point mass and assume a constant
ratio of sound speed to orbital speed (at a given phase in the spiral), 
as is required by strict self-similarity.
In our case, this ratio is not constant: it varies as
$r^{(1-\beta)/2}$ in general, and as $r^{1/4}$ in this section.
Nevertheless our models are self-similar when viewed in the limit
$c/v_{\rm orb}\to 0$, or equivalently $r^2\Sigma(r)/M\to 0$, and it
is only in this limit that they are exact.
We imagine that we are studying waves very close
to the central mass.

\subsection{Linear theory}

Much insight into the nonlinear dispersion relation can be obtained
from the linear limit, where
the surface density function (\ref{eq:spiraldef}) reduces to 
\begin{equation}
\Sigma(r,\theta)= S_0 r^{-3/2}[1+\epsilon\cos(\theta+\alpha\ln r)],
~~~~\Sigma_0(r) +\epsilon\Sigma_1(r,\theta),
\label{eq:linspiral}
\end{equation}
with $S_0$ a constant and $\epsilon\ll 1$.
The potential corresponding to (\ref{eq:linspiral}) is  [\pcite{Kalnajs}]
\begin{equation}
\Phi(r,\theta) = -2\pi G S_0 r^{-1/2}\left[K(0,0)
+\epsilon K(\alpha,1)\cos(\theta+\alpha\ln r)\right],
~~~~\equiv \Phi_0(r) +\epsilon\Phi_1(r,\theta),
\end{equation}
where
\begin{equation}
K(\alpha,m)= \left.\frac{1}{2}~
\Gamma\left(\frac{2m+1+2i\alpha}{4}\right)
\Gamma\left(\frac{2m+1-2i\alpha}{4}\right)\right/
\left[\Gamma\left(\frac{2m+3+2i\alpha}{4}\right)
\Gamma\left(\frac{2m+3-2i\alpha}{4}\right)\right].
\label{eq:Kalnajs}
\end{equation}
Linearization of the standard inviscid equations of motion 
yields the following dispersion relation:
\begin{equation}\label{eq:logdisp}
\kappa^2-(m\Omega-\omega)^2 
-\frac{2\pi G\Sigma_0(r)}{r}K(\alpha,m)\left(\alpha^2
+\frac{9}{4}\right)+\frac{c^2(r)}{r^2}\left(\alpha^2+\frac{9}{4}\right)=0.
\end{equation}
In the tight-winding limit $\alpha\gg m\ge 1$,
$K(\alpha,m)\approx|\alpha|^{-1}$, so that eq.~(\ref{eq:linWKBdisp})
is recovered.
Allowing for the effects of the effects of pressure on the equilibrium
rotation curve,
\begin{equation}
\frac{(\kappa^2-\Omega^2)r}{\pi G\Sigma_0(r)}=
\frac{1}{2}[K(0,0)-3\sigma^2],
\label{eq:ssfreqdiff}
\end{equation}
(cf. eq.~[\ref{gdef}]),
so that the dispersion relation for $m=1$ and $\omega=0$ 
can be put in dimensionless form:
\begin{equation}
K(0,0)-3\sigma^2 + (4\alpha^2+9)[\sigma^2-K(\alpha,1)]=0.
\label{eq:ssdimdisp}
\end{equation}

For any choice of $\sigma^2$, one might expect
two positive roots for $\alpha$, corresponding to the intersection
of the long and short branches with $\omega=0$.\footnote{\scite{LBO}'s 
Antispiral Theorem requires that all roots occur in pairs 
$\pm\alpha$: every
trailing spiral ($\alpha>0$) has a leading counterpart ($\alpha<0$).
We count only positive roots.}
However, equation~(\ref{eq:ssdimdisp}) has at most one root.
Whereas the tight-winding equivalent eq.~(\ref{lindisp}) is
positive as $kr\to0$,
the left hand side of eq.~(\ref{eq:ssdimdisp})
remains negative as $\alpha\to0$ because of the term
involving $K(\alpha,1)$.
Self-gravity is apparently too strong to allow the long-branch
wave to be stationary (see below).

The dispersion relation
has no real roots at all if $\sigma^2\ga 0.64136$.
At least for $\beta=1/2$ and $\gamma=4/3$, the maximum
Jean's length that permits these single-armed stationary waves to exist
is \begin{equation}\label{maxJean}
\left(\frac{\lambda_J}{r}\right)_{\max}\equiv 2\pi(\sigma^2)_{\max}= 4.030.
\end{equation}

It is clear that the real root for $\alpha$, where it exists,
belongs to the short branch.
Since the long-branch wave depends primarily on self-gravity,
its wavenumber should be independent of $\sigma^2\propto c^2$
as $c^2\to 0$,
yet eq.~(\ref{eq:ssdimdisp}) dictates that $\alpha\approx\sigma^{-2}$
as $\sigma^2\to0$.  [$K(\alpha,m)\approx |\alpha|^{-1/2}$ for
$\alpha\gg1$.]
However, the wave is ``short'' only in a relative sense.
For $m\ne1$ and except near resonances, the short wave normally 
has a wavelength comparable to the disk thickness or smaller.
In this case, the typical wavelength is $\sim\lambda_J$, which
is large compared to the disk thickness in the limit that
$M_{\rm disk}\ll M_{\rm star}$ (or equivalently $Q\to\infty$)
at fixed $\sigma^2$.
For our purposes, the thin-disk approximation is almost always appropriate.

To recap, \emph{there is no stationary long wave}.
However, there exist long waves with prograde pattern speeds.
Equation (\ref{eq:ssfreqdiff}) shows that
$\kappa^2-\Omega^2\propto r^{-5/2}$, whereas $\Omega\propto r^{-3/2}$.
To allow a long wave at small $\alpha$, the dispersion relation
(\ref{eq:logdisp}) requires $\omega>0$ and $\omega\propto r^{-1}$.
As $r\to 0$, $\omega$ is asymptotically negligible compared to
$\Omega(r)$ but asymptotically infinite
compared to the orbital frequency of any perturbing mass
that might launch waves inward.

\subsection{Nonlinear dispersion relation}

In the limit that the self-gravity and pressure of the disk are
very weak, every
fluid element must follow an orbit compatible with the dominant
potential of the central mass.
Hence the streamlines must be keplerian ellipses, although
their eccentricity need not be small.
The pressure and self-gravity of the disk control the choice
of the eccentricity and spirality $(e,\alpha)$ in a manner
described by the dispersion relation.
Since our problem is no longer local in radius, we cannot use
the elegant wave-action methods exploited in  Section 2 and instead
must look elsewhere for a physical condition to provide the
dispersion relation.

Note first of all that the net torque per unit mass on each streamline
must vanish, since the orbits are closed \footnote{Shocks
and other forms of dissipation could allow the gas to absorb
negative angular momentum from the waves, and then there would be
a nonzero torque balanced by accretion.}
But this condition does not provide the dispersion relation, because
it can be shown that \emph{the net torques on a streamline due to
pressure and gravity vanish separately for all choices of $e$ and $\alpha$}.
This is a property of logarithmic spirals for which the
surface density varies as $r^{-3/2}$ and the sound speed as $r^{-1/2}$
along a wave crest.
It is closely related to the fact that the angular momentum flux carried 
by such spirals is independent of radius [cf. \scite{ST96} and our
Appendix B].

The actual condition from which we derive the dispersion relation is that
the streamlines should not precess.
Streamlines become free-particle trajectories if one incorporates
the pressure and self-gravity of the disk into an effective potential,
\begin{equation}
V(r,\theta)= -\frac{GM}{r} + \Phi(r,\theta) + H(r,\theta),
\label{eq:Veff}
\end{equation}
where the first term on the right is due to the central
mass, the second term is the gravitational potential of the disk,
and the third term is the enthalpy,
\begin{equation}
H\equiv \frac{c^2}{\gamma-1},
\label{eq:enthalpydef}
\end{equation}
which varies with the surface density as $\Sigma^{\gamma-1}$.
(The adiabatic exponent appears in symbolic form for the sake
of clarity, but $\gamma\to 4/3$.)
The accelerations due to pressure are $-\grad H$.

Let the time average of $V$ along a keplerian ellipse be
$\overline{V}(a,e,\phi)$, where $\phi$ is the periapse angle.
Then by standard methods of secular perturbation theory,
the condition that the orbit not precess is (cf. \pcite{BC})
\begin{equation}
\frac{\partial\overline{V}}{\partial e}(a,e,\phi)\equiv 0,
\label{eq:noprecess}
\end{equation}
to which the keplerian part does not contribute.

It is convenient to introduce the eccentric anomaly $E$
and the true anomaly $\psi$.
The relation between the two anomalies and the time
in a keplerian orbit is (cf. \pcite{BC})
\begin{equation}\label{true_anom}
E-e\sin E=\psi\equiv n(t-t_{\rm peri}).
\end{equation}
Here $n=2\pi(a^3/GM)^{1/2}$ is the mean motion, and
$t_{\rm peri}$ is the time of pericenter.
If $\phi$ is the azimuth of pericenter, the position in
polar coordinates $(r,\theta)$ is 
\begin{eqnarray}
r&=& a(1-e\cos E),\nonumber\\
r\cos[\theta-\phi(a)] &=& a(\cos E-e),\nonumber\\
r\sin[\theta-\phi(a)] &=& a\sqrt{1-e^2}\sin E.
\label{eq:eccanom}
\end{eqnarray}
Because of self-similarity, the disk potential and enthalpy 
take the forms
\begin{equation}
\Phi(a,E)= a^{-1/2}f(E), ~~~~~H(a,E)= a^{-1/2}h(E),
\label{eq:streampot}
\end{equation}
where $f(E)$ and $h(E)$ are periodic functions.
It is sufficient to calculate $f$ and $h$ along a single streamline.

One can calculate $f(E)$ as
a double integral over all other streamlines:
\begin{eqnarray}
f(E)&=& -a^{1/2}G\int\limits_{a'=0}^{\infty} da'~a'\Sigma_0(a')
\int\limits_0^{2\pi} dE'\frac{1-e\cos E'}{R},\nonumber\\
R^2 &\equiv& a^2(1-e\cos E)^2 + a'^2(1-e\cos E')^2
- 2aa'\left\{ \left[(\cos E'-e)(\cos E-e) \vphantom{\sqrt{1-e^2}}
+(1-e^2)\sin E\sin E'\right]\cos(\alpha\ln a')\right.\nonumber\\
&& \phantom{a^2(1-e\cos E)^2 + a'^2(1-e\cos E')^2 -2~}
\left. -\left[(\cos E'-e')\sin E-(\cos E-e)\sin E'\right]\sqrt{1-e^2}
\sin(\alpha\ln a')\right\}.
\label{eq:gravint}
\end{eqnarray}
Here $R$ is the distance between the points corresponding to 
$E$ and $E'$ on the streamlines $a$ and $a'$, respectively.
The factor $(1-e\cos E')$ is proportional to the time spent in the
interval $dE'$, and
$\Sigma_0(a)$ is a lagrangian surface density defined so that
the mass between streamlines $a$ and $a+da$ as 
$2\pi\Sigma_0(a) ada =2\pi\Sigma_0(1)a^{-1/2}da$.

A  more efficient way to calculate the gravitational potential
is to decompose the surface density 
$\Sigma(r,\theta)$ into a sum of azimuthal Fourier harmonics,
each itself a logarithmic spiral, and then to use
Kalnajs' formula to obtain the corresponding harmonics of
$\Phi(r,\theta)$:
\begin{equation}
\Sigma(r,\theta)= r^{-3/2}
\sum\limits_{m=0}^\infty\hat\Sigma(m)
\cos m(\theta+\alpha\ln r),~~~~
\Phi(r,\theta)= -2\pi G r^{-1/2}\sum\limits_{m=0}^\infty
K(m\alpha,m)\hat\Sigma(m) \cos m(\theta+\alpha\ln r).
\label{eq:lsdecomp}
\end{equation}
Since $\Sigma(r,\theta)$ has the form (\ref{eq:spiraldef}),
the angular Fourier transform needed to find the coefficients 
$\hat\Sigma(m)$ need be performed only at one radius.
But we did not use this method except as a check.

The enthalpy and surface density are easily
calculated in streamline coordinates.
The transformation (\ref{eq:eccanom}) from
$(a,\psi)\to (r,\theta)$, when $e$ is fixed and $\phi(a)=-\alpha\ln a$,
has the Jacobian 
\begin{equation}
{\cal J}\equiv 
\frac{rdr d\theta}{adad\psi}
= \sqrt{1-e^2}+\alpha e\sin E.
\label{eq:Jacobian}
\end{equation}
Since
\begin{displaymath}
\frac{d\psi}{dE}=1-e\cos E=\frac{r}{a},
\end{displaymath}
we also have $drd\theta={\cal J}dadE$.
If the wave is an adiabatic deformation of a circular disk, the gas is
uniformly distributed with respect to
orbital phas $\psi$, so that $\Sigma r dr d\theta= \Sigma_0 a da d\psi$,
whence
\begin{equation}
\Sigma(a,E)= \Sigma_0(a){\cal J}^{-1},~~~~
H(a,E)= \sigma^2\cdot 2\pi G\Sigma_0(a)a {\cal J}^{1-\gamma}.
\label{eq:sigandh}
\end{equation}
Now $\cal J$ has zeros and $\Sigma$ has poles unless the
streamline-crossing parameter
\begin{equation}
q\equiv \alpha e/\sqrt{1-e^2}
\label{qdef}
\end{equation}
is less than unity.
So $q$ replaces the quantity $q_0\equiv kae$ of the tightly-wrapped theory.

The machinery above allows one to calculate the averages of
$\Phi$, $H$, and hence $V$ around the actual streamlines.
For example,
\begin{equation}
\overline{\Phi}= \int\limits_0^{2\pi}\Phi\,\frac{d\psi}{2\pi}=
\int\limits_0^{2\pi}\frac{dE}{2\pi}
(1-e\cos E)\Phi(a,E).
\end{equation}
But the streamline on which the average above is taken has the same
shape as all of the others, whereas to compute the variation
(\ref{eq:noprecess}), one needs to vary the eccentricity of the ``test''
streamline while holding all the rest---the ``field'' streamlines---fixed.
The potentials $\Phi$, $H$, and $V$ are regarded as fixed functions
of the Eulerian coordinates $(a_{\rm f},E_{\rm f})$ based on the
field streamlines, all of which have eccentricity $e_{\rm f}$
and spirality $\alpha=-d\phi_{\rm f}/d\ln a_f$: the subscripts
``$\rm f$'' and ``$\rm t$'' indicate field and test quantities, respectively.
A calculation yields
\begin{equation}
\left(\frac{\partial a_{\rm f}}{\partial e_{\rm t}}\right)_{a_{\rm t},
E_{\rm t},\phi_{\rm t}}= -\frac{a(e+\cos E)}{{\cal J}\sqrt{1-e^2}},~~~~
~~~~\left(\frac{\partial E_{\rm f}}{\partial e_{\rm t}}\right)_{a_{\rm t},
E_{\rm t},\phi_{\rm t}}=\frac{\sin E-\alpha\sqrt{1-e^2}\cos E}
{{\cal J}\sqrt{1-e^2}},
\end{equation}
in which $(a_{\rm t,f},E_{\rm t,f},\phi_{\rm t,f})\to
(a,E,\phi(a))$ after the differentiation.  Hence
\begin{equation}
\frac{\partial\overline{\Phi}}{\partial e}
= a^{-1/2}\int\limits_0^{2\pi}\frac{dE}{2\pi}
\left[\frac{1}{2}\left(e+\cos E\right) f(E) 
+\left(\sin E-\alpha\sqrt{1-e^2}\cos E\right)\frac{df}{dE}\right]
\frac{1-e\cos E}{{\cal J}\sqrt{1-e^2}}.
\label{eq:phivar}
\end{equation}
The precession rate due to pressure is given by an expression
identical to the above except that $\overline{H}~\&~h$
replace $\overline{\Phi}~\&f~$.

The results for $\sigma^2=0.2$ are shown by the points in Fig. 1.
We have checked the results for $e\ll 1$ against the linear
theory of  Section 3.1: $\alpha\approx 3.794$ (linear);
$\alpha\approx 3.792$ at $e=0.01$ (nonlinear).
The dispersion relation predicted
by the tight-winding theory of  Section 2 is also included
in Figure 1 using $q_0\equiv kae$ rather than $q$ as the abscissa.
Both the present self-similar solutions and the tight-winding
theory of  Section 2 predict that 
the spirals unwind with increasing nonlinearity
($d\alpha/dq<0$).
But the long branch predicted by the tight-winding approximation
does not exist.
Notice that like the tight-winding dispersion relation,
the self-similar sequence reaches a maximum $q$.
The linear analysis of  Section 3.1 shows, however, that the self-similar
sequence cannot return to $q=0=e$ at nonzero wavenumber $\alpha$.
Therefore it must terminate in a point at $\alpha=0$, $q>0$, and $e=1$.
The last (lowest) self-similar model shown
has $e=0.998$, $q=0.9016$, and $\alpha=0.057$, which corresponds
to a pitch angle of $86.7^\circ$.
Our numerical methods are unable to follow the sequence beyond this
point.

\subsection{Angular momentum flux}

There is more than one way to define the angular momentum flux of a
spiral wave.  Lynden-Bell \& Kalnajs (1972, henceforth LBK) 
imagine dividing the disk in
two by a cylinder coaxial with the rotation axis and consider the
total torque exerted by the interior of the cylinder on its exterior.
In collisionless disks, LBK's torque consists of two parts: a
gravitational stress that is quadratic in gradients of the potential,
and a ``lorry transport'' term that involves correlations between the
radial and azimuthal velocities of stars crossing the cylinder.  LBK's
angular momentum flux is Eulerian---it describes
the transfer of angular momentum from one spatial region to another.

Lagrangian approaches define the flux by the rate of transfer of
angular momentum from one mass element to another.  Lagrangian fluxes
arise naturally via Noether's theorem from an action principle in
which the dynamical variables follow the mass, but can also be defined
directly.

For our stationary spiral waves, we define the angular momentum
flux to be \emph{the total torque exerted by the disk interior to a
given streamline on the exterior.}  No ``lorry transport'' term occurs
because the gas does not cross streamlines, but the torque consists of
two parts, a gravitational term and a pressure term.  Appendix C shows
that these contributions to the flux can be obtained conveniently by
differentiating the internal and gravitational energies of the disk
with respect to radial wavenumber $\alpha$.\footnote{The total
internal and gravitational energies of the disk are infinite, but we
use the energies per unit logarithmic interval in semimajor axis,
which are finite.}  Thus the prescription for the flux is the same as
in the tight-winding theory; however, the gravitational flux so
obtained is strictly accurate only for exact logarithmic spirals in
disks where $\Sigma_0(a)\propto a^{-3/2}$.

For as much of the stationary sequence as we have computed,
the total angular momentum flux of the trailing self-similar spirals is
always positive and increases monotonically with streamline
eccentricity (Fig.~2).

\section{Summary and Discussion}
We have analysed stationary single-armed spiral density waves
in disks whose rotation curve is strongly dominated by the potential
of a pointlike central mass.  Both linear and nonlinear waves have
been considered.
For tightly-wound waves, we have
applied a phase-averaged variational formalism to derive 
the nonlinear dispersion relation, angular momentum density and flux,
and characteristic velocities of these waves
 ( Section 2 and Appendices A \& B).
The variational approach may be useful for other 
nonlinear dispersive waves of astrophysical interest \cite{Whitham}.
Unfortunately, when the waves are strongly nonlinear or when the
disk is barely cool enough to permit their existence
[eqs.~(\ref{sigmadef}) \& (\ref{maxJean})], the pitch angle is too
large for the tightwinding approximation.
Therefore we have constructed radially self-similar logarithmic spiral
solutions without restriction on the pitch angle, but only for disks
with azimuthally-averaged surface-density profiles scaling
as $r^{-3/2}$ ( Section 3).
Where both are valid, the self-similar and tight-winding
theories agree well.
The most important lessons learned are as follows
(unless otherwise noted, these statements apply to stationary, 
single-armed waves only):

\begin{enumerate}

\item Even though self-gravity is essential, the Toomre $Q$ parameter
is irrelevant to these waves.  This is because $Q$ depends upon
the central mass (via the orbital or epicyclic frequency), whereas
the existence---and even the wavelength or pitch angle---of these waves
is asymptotically independent of the mass ratio 
$M_{\rm central}/M_{\rm disk}$, since the waves consist of streamlines
approximating keplerian ellipses.
The appropriate dimensionless ratio of temperature
to self-gravity is not $Q^2$ but rather $\sigma^2$ [eq.~(\ref{sigmadef})],
which is equivalent to the ratio of Jeans length to radius.
A corollary is that the waves can propagate arbitrarily close to
the central mass provided $\sigma^2$ remains below its critical value
(\ref{maxJean}).

\item These waves are nonlinear extensions of the short branch
of the WKB dispersion relation (at zero pattern speed and azimuthal
wavenumber $m=1$); the long branch does not exist, at least not in
isentropic disks with surface densities $\Sigma\propto r^{-3/2}$
( Section 3.1).
Consequently, trailing waves propagate inward, even nonlinearly
(Appendix B).  Of course the long wave does exist at other arm
multiplicities or retrograde pattern speeds.

\item If the azimuthally-averaged surface density profile
is shallower (steeper) than $r^{-3/2}$, then ingoing
(outgoing) waves become increasingly nonlinear as they propagate.
This is a rather general result; it follows from the fact that
the gravitational angular momentum flux scales
as $G\Sigma^2 r^3$ at fixed pitch angle and eccentricity.

\end{enumerate}

Regarding the first point above, it is always easier to 
have $\sigma^2 < 0.64$
in a keplerian disk than to have $Q\la 1$ since
\begin{equation}\label{relation}
\sigma^2= \frac{H}{2r}Q,
\end{equation}
where $H\equiv v_s/\Omega$ is the disk thickness.
Of course the single-armed waves we have
studied are not locally unstable modes.
On the other hand, it is interesting that the
Jeans length $\lambda_J=2\pi r\sigma^2$ is approximately
the wavelength of fastest growth 
for \emph{axisymmetric secular} instabilities
(\pcite{LBP}, \pcite{FP84}).
Viscous effects violate the conservation
of vorticity, which tends to stabilize the disk at long wavelengths.
Secular instability does not actually occur at large $Q$, however, 
unless the viscous torque is a constant or decreasing function of
surface density \cite{ST:95}.

With regard to the second point, it should be noted that
although the waves we study belong formally to the short branch,
their wavelengths are nevertheless long compared to the disk
thickness when $Q\gg1$.
In fact their characteristic wavelength is $\lambda_J=\pi Q H$.

It is interesting that the minimum mass solar nebula, defined
by augmenting the present planets with enough hydrogen and helium
to reach solar abundance, has an $r^{-3/2}$ surface density
profile.
For example, \pcite{Hayashi} quote
\begin{equation}\label{minneb}
\Sigma(r)= 1.7\times 10^3\left(\frac{r}{1~\AU}\right)^{-3/2}
~\mbox{g cm}^{-3},~~~~~
T(r)= 280\left(\frac{L_*}{L_\odot}\right)^{1/4}
\left(\frac{r}{1~\AU}\right)^{-1/2}.
\end{equation}
Very possibly this is only a coincidence.  On the other hand,
if density waves of the sort we have analysed were to dominate
angular momentum transport, they might drive the
surface density towards this power law: for a shallower
profile, ingoing trailing waves would tend to steepen and
dissipate at small radii, depositing negative angular momentum
and driving accretion.
The temperature profile above is shallower than expected
for a thin optically thick disk warmed either by accretion
or by reprocessing of solar radiation ($T\propto r^{-3/4}$);
however, the $r^{-1/2}$ scaling seems required to match integrated
infrared spectra \cite{ALS} and may result from flaring of the disk
(\pcite{KH}; \pcite{CG}).  For the profiles (\ref{minneb}),
the parameter $\sigma^2$ is independent of radius:
$\sigma^2\approx 1.5(L_*/L_\odot)^{1/4}$.  
This is comparable to the critical value $0.64$.  Under the same conditions,
$Q\approx 70 (r/\AU)^{-1/4}$.

VLBI observations of extragalactic $\rm H_2O$ water masers have revealed
molecular disks in the nuclei of NGC 4258 \cite{Miyoshi} and
NGC 1068 \cite{GGAB} on subparsec scales.
In the former case, the disk appears to be very thin, though
warped, and the rotation accurately keplerian, being dominated
by a $3.6\times 10^7~M_\odot$ black hole.
If one assumes that the observed Xray luminosity of the nucleus,
$L_X= 4\times 10^{40}~\mbox{erg~s}^{-1}$ \cite{Makishima},
is powered by \emph{steady} accretion through this disk, then
(cf. \pcite{NM95})
\begin{displaymath}
\sigma^2 \approx 7\times 10^{-3}\epsilon_{0.01}\alpha T_{300}^2
r_{0.1}^{3/2},~~~~~
Q\approx 14.\epsilon_{0.01}\alpha T_{300}^{3/2}r_{0.1},
\end{displaymath}
where $100\epsilon_{0.01}$ is the efficiency with which mass is converted
to Xrays, $\alpha\la 1$ is the usual viscosity parameter,
$T_{300}$ is the midplane temperature in units of 300~K (the minimum
for masing), and $10 r_{0.1}~\mbox{pc}$ is the distance from the black hole.
Given the uncertainty in the first three parameters, it could easily
be that $Q\la 1$, but in any case the single-armed waves are
surely permitted; the question is whether there is any way to excite them.

\section*{Acknowledgments}
We thank Scott Tremaine 
and Steve Lubow for helpful discussions, and the Isaac
Newton Institute for Mathematical Sciences for its hospitality
to JG while some of this work was carried out.
This research was supported by NASA Astrophysical
Theory Grant NAG5-2796.

\begin{appendix}
\section{Averages over orbital phase for tightly-wound waves}

After substitution from eq.~(\ref{delta_r}), the phase average of
the internal energy (\ref{W_int}) becomes
\begin{equation}
\bar W_{\rm int} = \int 2 \pi a da \frac{\Sigma c^2} {\gamma (\gamma-1)}
\int_0^{2 \pi} \frac {d \Psi} {2 \pi} \left[ \left(1-q_0 \sin \Psi \right)
^{-(\gamma-1)} -1 \right]~~~
=\int 2 \pi a da \left[\frac{\Sigma c^2}{4}\, q_0^2 {\cal U}(q_0)\right].
\end{equation}
The phase-averaged expression is independent of $\theta$, integration
over which has yielded the factor of $2\pi$.
The definition (\ref{Udef}) of ${\cal U}(q_0)$ is closely related
to an integral representation of the hypergeometric function
\cite{AS}, whence
\begin{equation}
{\cal U}(q_0^2)= \frac{4}{\gamma(\gamma-1)q_0^2}\left[
\vphantom{F}_2F_1\left(\frac{\gamma-1}{2},\frac{\gamma}{2},1;q_0^2\right)
~-1\right].
\end{equation}

In the gravitational integral (\ref{W_grav}), we
replace the limits on $\Delta a\equiv a_1-a_2$ with $\pm\infty$
(a good approximation for tightly-wound waves) and eliminate
$\theta$ in favor of $\Psi$ using (\ref{kepler_phase}):
\begin {equation}
W_{\rm grav}= 2 G \int a \Sigma^2(a) da \int_0^{\infty} d \Delta a
 \int_0^{2\pi}d\Psi
\ln \left|1+\frac {ae\cos(\Psi+k\Delta a/2)-ae\cos(\Psi-k\Delta a/2)} 
{ \Delta a} \right| 
\end {equation} 
The integration over $\Psi$ can be cast in the form
\begin{equation}\label{Fint}
\int_0^{2 \pi} \frac {d \Psi} {2 \pi} \ln \left|1-
R \sin \Psi \right|~=~\ln \left|\frac 1 2  + \frac 1 2 \sqrt{1-R^2}\right|
\end{equation}
where $R\equiv 2ae\sin(k\Delta a)/\Delta a$, or equivalently
$R= q_0\sin(q_0 y)/y$ if $y\equiv \Delta a/2ae$.
The integration over $\Delta a$ can then be expressed in terms of
the function ${\cal W}(q_0)$ defined in eq.~(\ref{Wdef}).
The gravitational energy is therefore
\begin {equation}
\bar W_{\rm grav} = 
\int 2 \pi a da \left[
-\frac {\pi G \Sigma^2}{2} ae |q_0| {\cal W}(q_0)\right].
\end {equation}
We have not been able to relate ${\cal W}(q_0)$ to
familiar special functions, so we evaluate it by expanding the
the integrand of eq.~(\ref{Wdef}) in powers of $q_0^2$ and
integrating term by term.

\section {Nonlinear Characteristic Velocities}

The characteristic velocities of a nonlinear wave train
represent the speeds of propagation for modulations of
the amplitude, wavelength, and frequency.
Their derivation is simplified in our case by
the special structure of the action (\ref{Iavg})-(\ref{Lbar})
in the nearly stationary limit:
\begin{eqnarray}\label{slow_action}
I&=& \int\int\left[{\cal L}_0(A,k,a)-\omega A\right]J(a)da dt~+O(\omega^2),\\
{\cal L}_0 &\equiv&
-\frac {\pi G\Sigma_0 a}{2a\Omega} \left[\frac 1 2 g e^2 
- |q_0| {\cal W}(q_0) e + \sigma^2 q_0^2 {\cal U}(q_0) \right],
\label{Lcurly}
\end{eqnarray}
where $A\equiv e^2/2$ and $J(a)\equiv 2\pi a^3\Sigma_0(a)\Omega(a)$.
The $O(\omega^2)$ terms do not
not contribute to the characteristic velocities as
$\omega\to 0$ and will be neglected.
Variation of $I$ with respect to $A$ yields the dispersion relation
\begin{equation}\label{disp_formal}
\omega(A,k,a)= \frac{\partial{\cal L}_0}{\partial A},
\end{equation}
and variation with respect to the phase function $\Psi$
yields the conservation of wave action (cf. eq.~[\ref{Euler_Lagrange}])
\begin{equation}\label{EL_formal}
\frac{\partial}{\partial t}\left(JA\right) 
+\frac{\partial}{\partial a}\left(J\frac{\partial{\cal L}_0}{\partial k}
\right)=0.
\end{equation}
On the other hand, because
partial derivatives of $\Psi$ commute, it follows from 
eqs.~(\ref{dPsidef}) that
\begin {equation}\label{consistency}
\frac {\partial k} {\partial t} +  \frac {\partial\omega}{\partial a}= 0.
\end {equation}

The derivatives with respect to $a$ in
equations (\ref{EL_formal}) and (\ref{consistency})
include the indirect dependence of ${\cal L}_0$ and $\omega$
on $a$ via their dependence on $k$ and $A$.
Substituting from eq.~(\ref{disp_formal}) and expanding derivatives,
we have 
\begin{eqnarray}\label{vsystem}
\frac {\partial k} {\partial t} ~+~ 
\left(\frac{\partial^2{\cal L}_0}{\partial A\partial k}\right)
\frac{\partial k}{\partial a}~+~
\left(\frac{\partial^2{\cal L}_0}{\partial A^2}\right)
\frac{\partial A}{\partial a} &=&
-\frac{\partial{\cal L}_0}{\partial a}\nonumber\\
\frac {\partial A} {\partial t} ~+~ 
\left(\frac{\partial^2{\cal L}_0}{\partial k^2}\right)
\frac{\partial k}{\partial a}
~~+~\left(\frac{\partial^2{\cal L}_0}{\partial k\partial A}\right)
\frac{\partial A}{\partial a}
&=& -\frac{\partial^2{\cal L}_0}{\partial a\partial k}
-\frac{\partial{\cal L}_0}{\partial a}\frac{\partial\ln J}{\partial a},
\end{eqnarray}
where now $\partial {\cal L}_0/\partial a$ involves only
the explicit dependence of ${\cal L}_0(A,k,a)$ on its third argument.

Equations (\ref{vsystem}) form a system of first-order quasilinear
partial differential equations for $(k,A)$ in the $(a,t)$ plane.
Such systems can be solved by integrating along characteristic curves
$da/dt= V(k,A,a,t)$.
The characteristic velocities $V$ are eigenvalues of the 
matrix of coefficients
of $\left(\frac{\partial k}{\partial a},\frac{\partial A}{\partial a}
\right)$:
\begin{equation}\label{V_formal}
V_1,V_2 = \frac{\partial^2{\cal L}_0}{\partial k\partial A}
~\pm~\sqrt{\frac{\partial^2{\cal L}_0}{\partial A^2}
\frac{\partial^2{\cal L}_0}{\partial k^2} }~~~
=~~ \frac{\partial\omega}{\partial k}
~\pm~\sqrt{\frac{\partial\omega}{\partial A}
\frac{\partial^2{\cal L}_0}{\partial k^2} }~.
\end{equation}
The reduced lagrangian (\ref{Lcurly}) has a Taylor series in $A$ beginning
at $O(A)$, and $\omega$ has a Taylor series beginning at $O(1)$.
Therefore as $A\to 0$, the characteristic velocities merge and can
be identified with the linear group velocity,
but at finite amplitude they are distinguished by
corrections of order $A^{1/2}=O(e)$.
Should the argument of the surd in (\ref{V_formal}) be negative, the
characteristic velocities are complex; this may signal
a tendency of the nonlinear wave train to break up into a series of
solitary waves \cite{Whitham}.
The required derivatives of ${\cal L}_0$ are
\begin{eqnarray}\label{secondderiv}
\frac{\partial^2{\cal L}_0}{\partial k\partial A}=
\frac {\partial \omega} {\partial k}&=& \frac {\pi G \Sigma} {2\Omega}
\left[-\sigma^2 ka \left(q_0^2{\cal U}^{\prime
\prime} + 5 q_0 {\cal U}^\prime + 4 {\cal U}\right) + 
\left(q_0^2 {\cal W}^{\prime \prime} + 4 q_0 {\cal W}^\prime
 + 2{\cal W}\right)\mbox{sign}(k) \right],\nonumber\\
\frac{\partial^2{\cal L}_0}{\partial A^2}=
\frac {\partial \omega}{\partial A} &=& \frac {\pi G \Sigma} {2\Omega}
\frac {k^2a} {e^2} \left[- \sigma^2 \left(q_0^2 {\cal
U}^{\prime \prime} +3 q_0 {\cal U}^\prime \right) 
+  e\left(q_0{\cal W}^{\prime
\prime} + 3 {\cal W}^\prime \right)\mbox{sign}(k) \right],\nonumber\\
\frac {\partial^2{\cal L}_0} {\partial k^2}
\phantom{~=\frac {\partial \omega}{\partial A}}
 &=& \frac {\pi G \Sigma}
{2\Omega} a e^2 \left[- \sigma^2\left( q_0^2{\cal
U}^{\prime \prime} + 4 q_0 {\cal U}^\prime + 2 {\cal U} \right) + 
e\left(q_0 {\cal W}^{\prime \prime} + 
2 {\cal W}^\prime\right)\mbox{sign}(k)  \right].
\end{eqnarray}

The characteristic velocities are shown for representative values
of $\sigma^2$ in Figure~\ref{velfig}.
We have plotted results for the short branch only, since the the
tight-winding approximation is not applicable to the long branch.
In all cases that we have examined, both chararacteristic
velocities are real and negative for tightly wrapped trailing waves
($ka\gg1$, $k>0$.)

\section{Angular momentum flux of self-similar spirals}

We relate the angular momentum flux of nonlinear logarithmic
spirals to derivatives of the gravitational and
internal energies with respect to pitch angle.
The flux proves to be constant in isentropic disks with
$\Sigma_0(a)\propto a^{-3/2}$ and adiabatic index $\gamma=4/3$,
so that no net torque is exerted on streamlines.

Consider first the gravitational interaction energy between
two keplerian elliptical rings of the same eccentricity $e$,
semimajor axes $(a_1,a_2)$, periapse angles $\phi_1,\phi_2$, and
masses $(m_1,m_2)$.
After integration over the arc lengths, the
energy can be put into the form [cf. eqs~(\ref{eq:eccanom}) \&
(\ref{eq:gravint})]
\begin{equation}
W_{12} = -\left(\frac{Gm_1m_2}{\sqrt{a_1a_2}}\right)
 K\left(\ln\frac{a_2}{a_1},~\phi_2-\phi_1\right).
\label{eq:tworing}
\end{equation}
Because we have factored out $(a_1a_2)^{-1/2}$,
the dimensionless kernel $K(\xi,\phi)$ is independent of the 
absolute dimensions of the rings, and it is symmetric in the sign
of both arguments.
The torque exerted by ring 2 on ring 1 is, with $x\equiv\ln a$,
\begin{equation}
\Gamma_{12}=-\frac{\partial W_{12}}{\partial\phi_1}=
-\frac{Gm_1m_2}{\sqrt{a_1a_2}} 
K_\phi(x_2-x_1,\phi_2-\phi_1),
\label{eq:G12}
\end{equation}
where $K_\phi(\xi,\phi)\equiv\partial K(\xi,\phi)/\partial\phi$.
In a continuous elliptical disk, let
\begin{equation}
\mu\equiv 2\pi a^{3/2}\Sigma_0(a),
\end{equation}
so that the mass between streamlines
$a$ and $a+da$ is $a^{1/2}\mu dx$, and
$\mu$ is constant since $\Sigma_0\propto a^{-3/2}$.
The gravitational energy per logarithmic interval in semimajor
axis is therefore
\begin{equation}
W(x_1,\phi_1) = -\frac{G\mu^2}{2}\int\limits_{-\infty}^\infty
K(x_2-x_1,\phi_2-\phi_1) dx_2,
\label{eq:W0}
\end{equation}
and the total torque per unit $x$ is
\begin{equation}
\frac{\partial W}{\partial\phi_1}(x_1,\phi_1) = 
-\frac{G\mu^2}{2}\int\limits_{-\infty}^\infty
K_\phi(x_2-x_1,\phi_2-\phi_1) dx_2.
\label{eq:T0}
\end{equation}
In a logarithmic spiral, $\phi_2-\phi_1=-\alpha\cdot(x_2-x_1)$.
Then the energy (\ref{eq:W0}) is independent of $x_1$.
However, the energy does depend upon the pitch angle:
\begin{equation}
\frac{\partial W}{\partial\alpha}=
G\mu^2\int\limits_{0}^\infty
\xi K_\phi(\xi,-\alpha\xi) d\xi.
\end{equation}

The total torque exerted by all rings $x_2<x$ on all rings $x_1>x$
does not vanish.
Referring to eq.~(\ref{eq:G12}), and noting the antisymmetry
of $K_\phi$ in its second argument, this torque is
\begin{eqnarray}
\Gamma_{\rm grav}(x,\alpha)&=& +G\mu^2\int\limits_x^{\infty} dx_1
\int\limits_{-\infty}^x dx_2 K_\phi[x_1-x_2,-\alpha(x_1-x_2)]
\nonumber\\
&=&G\mu^2\int\limits_0^\infty d\xi\int\limits_{x-\xi/2}^{x+\xi/2}
d\eta K_\phi(\xi,-\alpha\xi)
~~=~~G\mu^2\int\limits_0^\infty \xi K_\phi(\xi,-\alpha\xi)d\xi
~~=~~\frac{\partial W}{\partial\alpha}.
\label{eq:Gravflux}
\end{eqnarray}
This establishes the desired relationship between the gravitational
contribution to the angular momentum flux and the derivative of
the gravitational energy with respect to pitch angle.
Since $\partial\Gamma_{\rm grav}/\partial x=0$, there is no net torque
on streamlines.

Now consider the pressure term.
For a polytropic gas, the internal energy per unit mass is
\begin{displaymath}
\frac{p}{\gamma-1}=\frac{p_0}{\gamma-1}\left(\frac{V}{V_0}\right)^{-\gamma},
\end{displaymath}
where $p$ is the pressure, $V=1/\Sigma$ is the specific volume,
and the subscript ``0'' denotes a reference state 
on the gas adiabat.
Integrating around a streamline, and noting the area element
$a^2{\cal J}dEdx$, we have for the internal
energy per logarithmic interval of semimajor axis,
\begin{equation}
U(a,\alpha)= \frac{p_0(a)a^2}{\gamma-1}
\int\limits_0^{2\pi} {\cal J}^{1-\gamma} dE,
\end{equation}
where $E$ is the eccentric anomaly and ${\cal J}$ is the
Jacobian (\ref{eq:Jacobian}).
The derivative with respect to $\alpha$ is
\begin{equation}
\frac{\partial U}{\partial\alpha}=
p_0(a)a^2
\int\limits_0^{2\pi} e\sin E {\cal J}^{-\gamma} dE.
\label{eq:dUdalpha}
\end{equation}

The angular momentum flux due to pressure is
the integral of $r\ve_\theta\cdot\vT\cdot\vn=rp\ve_\theta\cdot\vn$
along the arc length of the streamline, where $\vT$ is the pressure tensor,
$\ve_\theta$ a unit azimuthal vector,
and $\vn$ the outward unit normal to the streamline.
If $a=a(r,\theta)$ then
$\vn=|\grad a|^{-1}\grad a$, and the area element can be written
$|\grad a|^{-1}ds da= {\cal J}a da dE$,
$ds$ being the element of arc length.
Hence,
\begin{equation}
\Gamma_{\rm press} = \int\limits_0^{2\pi} p(a,E) a {\cal J}
\left(\frac{\partial a}{\partial\theta}\right)_r dE.
\end{equation}
Now $\gamma p(a,E)=\Sigma_0(a)c_{s,0}^2(a){\cal J}^{-\gamma}$,
and equations (\ref{eq:eccanom}) yield
$(\partial a/\partial\theta)_r = -{\cal J}^{-1}a e\sin E$.
With these substitutions, we have
\begin{equation}
\Gamma_{\rm press}= \frac{\partial U}{\partial\alpha},
\end{equation}
as claimed.
Since the prefactor $p_0^2(a)a^2$ is independent
of $a$ in our self-similar spiral disks, $\Gamma_{\rm press}$
is also independent of $a$, so that the net pressure torque on
a streamline vanishes.

\end{appendix}
\bsp
\end{document}